\documentclass{JHEP3}
\usepackage{amsmath}
\input epsf
\usepackage{epsfig}
\usepackage{amssymb}
\usepackage[active]{srcltx}
\usepackage{graphicx}

\newcommand{\bea}{\begin{eqnarray}}
\newcommand{\eea}{\end{eqnarray}}
\newcommand{\bean}{\begin{eqnarray*}}
\newcommand{\eean}{\end{eqnarray*}}

\def\braket#1{\left\langle #1 \right\rangle}

\def\gb #1{ \left\langle #1 \right]}
\def\tgb #1{ \left[ #1 \right\rangle}

\def\a{{\alpha}}

\def\b{{\beta}}

\def\vev{\braket}
\def\tgb #1{ \left[ #1 \right\rangle}

\def\bvev#1{\left[ #1 \right]}
\def\Spaa{\vev}
\def\Spbb{\bvev}
\def\Spab{\gb}
\def\Spba{\tgb}

\def\Label#1{\label{#1}%
  \smash{\hbox to0pt{\raise1ex\hbox{\tiny[#1]}\hss}}}

\title{Tree amplitudes of noncommutative $U(N)$ Yang-Mills Theory}
\author{Jia-Hui Huang$^{\diamond}$, Rijun Huang$^{\dagger}$, Yin Jia$^{\dagger}$ ~~~~\\
$^\diamond$Center of Mathematical Science, Zhejiang University,
Hangzhou, China\\
$^{\dagger}$Zhejiang Institute of Modern Physics, Physics Department, Zhejiang University, Hangzhou, China\\
}
\date{\today}
\abstract{Following the spirit of S-matrix program, we proposed a
modified Britto-Cachazo-Feng-Witten recursion relation for tree
amplitudes of noncommutative $U(N)$ Yang-Mills theory. Starting from
three-point amplitudes, one can use this modified BCFW recursion
relation to compute or analyze color-ordered tree amplitudes without
relying on any detail information of noncommutative Yang-Mills
theory. After clarifying the color structure of noncommutative tree
amplitudes, we wrote down the noncommutative analogies of
$U(1)$-decoupling, Kleiss-Kuijf and Bern-Carrasco-Johansson
relations for color-ordered tree amplitudes, and proved them using
the modified BCFW recursion relation.  }

\begin{document}

\section{Introduction}

Quantum field theory has been proved to be an efficient way to
describe the world, yet one does still  not fully understand it.
Toward the understanding of quantum field theory, many approaches
have been suggested, and one of them is S-matrix
program\cite{S-matrix}. In the framework of S-matrix program, one
tries to study quantum field theory without relying on any detail
information but some general principles, such as Lorentz invariance,
Locality, Gange symmetry, etc. Due to the appearance of on-shell
BCFW recursion relation\cite{Britto:2004ap,Britto:2005fq}, it is now
possible and much easier to analyze or calculate scattering
amplitudes in the framework of S-matrix program. Together with
unitarity
method\cite{Bern:1994cg,Bern:1994zx,Britto:2004nc,Britto:2005ha},
one can compute one-loop amplitudes and tree amplitudes efficiently.

In conventional field theory with gauge symmetry, the full
scattering amplitude can be expressed as sum of color-ordered
amplitudes, known as color
decomposition\cite{Berends:1987cv,Mangano:1987xk,Mangano:1988kk,Dixon:1996wi}.
Thus in order to compute the full amplitude, one just need to
calculate color-ordered amplitudes, which are often much easier to
deal with. There are totally $(n-1)!$ color-ordered amplitudes for
$n$-point scattering amplitudes, but thanks to some relations of
amplitudes, we do not need to calculate all of these $(n-1)!$ ones.
There is so called $U(1)$-decoupling relation which states that
certain linear combination of partial amplitudes must vanish.
Besides this, there are also Kleiss-Kuijf
relations\cite{Kleiss:1988ne} which will reduce independent
amplitudes to $(n-2)!$ and Bern-Carrasco-Johansson
relations\cite{Bern:2008qj} which will further reduce independent
amplitudes to $(n-3)!$. Then it is possible to compute these
independent amplitudes starting from three-point amplitudes by BCFW
recursion relation, and three-point amplitudes can be determined
through the argument based on general principles of field theory and
the three-point
kinematics\cite{Benincasa:2007xk,ArkaniHamed:2008gz,Schuster:2008nh,He:2008nj}.
What is more, all the above mentioned relations of amplitudes can be
proved using BCFW recursion relation\footnote{The KK relations have
been proved by field theory method in \cite{DelDuca:1999rs}. Both KK
and BCJ relations have been proved by string theory method in
\cite{BjerrumBohr:2009rd,Stieberger:2009hq}. An extension of BCJ
relations to matter fields can be found in
\cite{Sondergaard:2009za}. See further works
\cite{Tye:2010dd,BjerrumBohr:2010zs,Mafra:2009bz}.
}\cite{Feng:2010my}(See further generalization and discussion
\cite{Jia:2010nz,Zhang:2010ve}). This again showed the power of BCFW
recursion relation in S-matrix program.

After these achievements of BCFW recursion relation in conventional
field theory, it is natural to think whether this powerful on-shell
recursion relation can be applied to nonlocal theories, one of which
we are interested in is noncommutative field
theory\cite{Douglas:1997fm,Cheung:1998nr,Chu:1998qz,Seiberg:1999vs}
(see also review papers \cite{Douglas:2001ba,Szabo:2001kg} and
references there in). Noncommutative field theory is a modification
of field theory obtained by taking the position coordinates to be
noncommutative variables, i.e., the coordinates satisfy
\bea [x^\mu,x^\nu]=i\theta^{\mu\nu}~,~~~\eea
where $\theta^{\mu\nu}$ is a constant antisymmetric tensor of
dimension (length)$^2$. This will in turn modify the Lagrangian of
field theory, and further bring modification to Feynman rules. It
has been shown that in noncommutative $U(N)$ Yang-Mills theory,
Feynman rules for propagators stay the same while those for vertices
are changed\cite{Armoni:2000xr,Bonora:2000ga}. Thus the analytic
structure of noncommutative amplitudes are very different from that
of conventional field theory. Besides of ordinary singularities from
propagators, there are also, even at the tree level, so called
essential singularities in complex plane\cite{Raju:2009yx}. These
essential singularities would disable the application of original
BCFW recursion relation in noncommutative field theory. Due to a
nice property that at tree level noncommutative amplitudes can be
expressed as ordinary amplitudes multiplied additional phase
factors\cite{Filk:1996dm,Minwalla:1999px,Martin:1999aq,Hayakawa:1999zf,Hayakawa:1999yt,Matusis:2000jf,Douglas:2001ba},
in \cite{Raju:2009yx} the author argued that one can obtain ordinary
amplitudes using BCFW recursion relation of conventional field
theory and then multiply them by corresponding phase factors to get
noncommutative amplitudes, thus avoid the effect of essential
singularities. In \cite{Boels:2010bv} the authors also argued that
BCFW recursion relation could be extended to noncommutative field
theory from the view of string theory. In this note, by removing
essential singularities from amplitudes of noncommutative $U(N)$
field theory, we introduce a modified BCFW recursion relation which
can be applied directly to noncommutative amplitudes. Using this
modified BCFW recursion relation, it is possible to construct any
tree amplitudes of noncommutative $U(N)$ Yang-Mills theory from
three-point noncommutative amplitudes.

Since BCFW recursion relation is often applied to color-ordered
amplitudes, we also investigate color structures of vertices of
noncommutative $U(N)$ Yang-Mills theory. The color algebra of
noncommutative $U(N)$ Yang-Mills theory has been discussed in many
papers, and Feynman rules for propagators and vertices have been
also worked out\cite{Douglas:2001ba,Armoni:2000xr,Bonora:2000ga}. By
working out color structures of vertices we show in detail how to
decompose the full noncommutative amplitude into color-ordered
noncommutative amplitudes. Based on the color structure, we are able
to discuss nontrivial relations among these amplitudes. It will be
shown that after modifications, $U(1)$-decoupling, KK and BCJ
relations can also be held for noncommutative amplitudes\footnote{In
\cite{Boels:2010bv}, by considering open string theory in a non-zero
constant $B$-field background\cite{Seiberg:1999vs}, some monodromy
relations are proposed. The field theory limits of these monodromy
relations are nontrivial KK and BCJ relations of noncommutative
amplitudes mentioned above. }, and all these relations can be proved
by modified BCFW recursion relation for noncommutative $U(N)$
Yang-Mills theory. We should emphasize that all these things, such
as computing tree level noncommutative amplitudes and proving
nontrivial relations, can be done in the framework of S-matrix
program, by using only noncommutative BCFW recursion relation and
three-point amplitudes. This beautifully illustrates the idea of
S-matrix program.

This note is organized as follows. In section two we will briefly
review the color algebra of noncommutative $U(N)$ Yang-Mills theory,
and discuss color structures of three-gluon and four-gluon vertices.
These lead to the color decomposition of noncommutative amplitudes.
We will also present two useful relations considering cyclic
permutation and reflection of color ordering. In section three we
discuss the validation of BCFW recursion relation in noncommutative
theory, and suggest one modified BCFW recursion relation for
noncommutative amplitudes. We also discuss three-point amplitudes
and show one simple example to verify BCFW recursion relation and
three-point amplitudes of noncommutative theory. In section four we
write down noncommutative analogies of $U(1)$-decoupling, KK and BCJ
relations and prove them through BCFW recursion relation of
noncommutative $U(N)$ Yang-Mills theory. In the last section some
general discussions and conclusions will be offered.

\section{The color structure}

\subsection{Color algebra and color structure of vertices}

Let us consider scattering amplitudes of noncommutative $U(N)$
Yang-Mills theory\cite{Douglas:2001ba,Armoni:2000xr,Bonora:2000ga}.
It is known that $U(N)$ group can be expressed as direct product of
$SU(N)$ and $U(1)$ groups. Let $t^a$ be the generator of $SU(N)$
group, where $a$ takes value from 1 to $N^2-1$, and $t^a$ is
normalized as
\bea \mbox{tr}(t^at^b)={1\over 2}\delta^{ab}~.~~~~\eea
Further more, these generators satisfy
\bea [t^a,t^b]=if^{abc}t^c~,~~~~\{t^a,t^b\}={1\over
N}\delta^{ab}+d^{abc}t^c~,~~~~\eea
thus we define the structure constant $f^{abc}$ and another
completely symmetric tensor $d^{abc}$. We also introduce $t^0$ as
the generator of $U(1)$ group, and for normalization we set
\bea t^0={1\over \sqrt{2N}}I_{N\times N}~.~~~\eea
These generators together lead to generators of $U(N)$ group. If we
denote $t^A$ as the generator of $U(N)$ group, where $A$ takes value
from 0 to $N^2-1$, then these generators satisfy normalization
condition
\bea \mbox{tr}(t^At^B)={1\over 2}\delta^{AB}~,~~~~\eea
and other two relations
\bea
[t^A,t^B]=if^{ABC}t^C~,~~~\{t^A,t^B\}=d^{ABC}t^C~.~~~\label{unalgebra}\eea
One should note that $SU(N)$ generators $t^a$ are traceless while
$U(N)$ generators $t^A$ do not strictly hold this property.

\begin{figure}[t]
\center
  \includegraphics[width=6in]{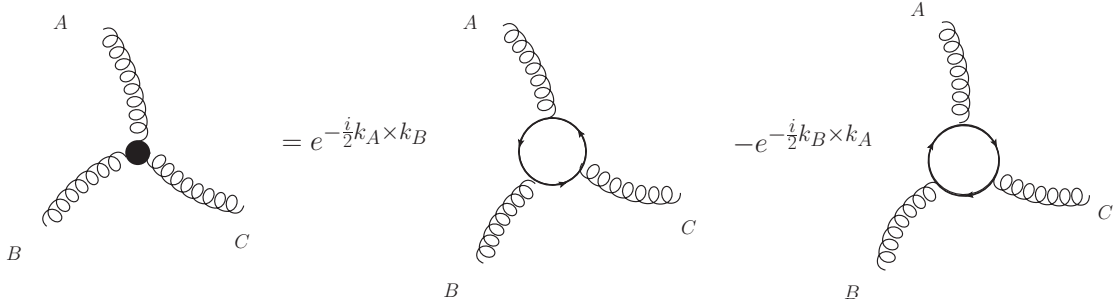}\\
  \caption{Color structure of three-gluon vertex of noncommutative $U(N)$
  Yang-Mills theory in double line notation. The order of generators in trace
strings are represented by the direction of arrows, and overall
factor has been stripped off.}\label{3pcolor}
\end{figure}

\begin{figure}[t]
\center
  \includegraphics[width=6.5in]{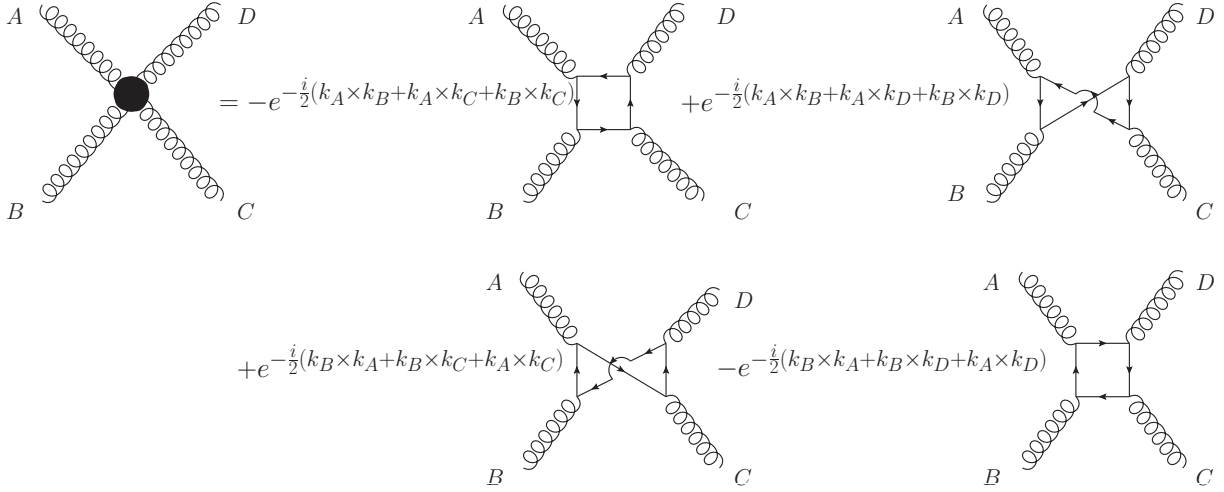}\\
  \caption{Color structure of four-gluon vertex of noncommutative $U(N)$
  Yang-Mills theory in double line notation. The order of generators in trace
strings are represented by the direction of arrows, and overall
factor has been stripped off.}\label{4pcolor}
\end{figure}

What is the difference between color structures of noncommutative
field theory and conventional field theory? This can be seen from
three-point and four-point vertices in both theories. In
conventional Yang-Mills theory only $f^{abc}$ is assigned to
three-point vertex, and from color algebra we
have\cite{Dixon:1996wi}
\bea f^{abc}\sim \mbox{tr}(t^at^bt^c)-\mbox{tr}(t^bt^at^c)~.~~~\eea
But in noncommutative Yang-Mills theory both $f^{ABC}$ and $d^{ABC}$
would present in one vertex. This difference will introduce
additional phase factors in color-ordered amplitudes of
noncommutative field theory. More concretely, we know that there are
three kinds of interaction for three-gluon vertex, namely
$SU(N)-SU(N)-SU(N)$, $SU(N)-SU(N)-U(1)$ and $U(1)-U(1)-U(1)$
interactions\cite{Armoni:2000xr}. All these three kinds of
interactions can be formally expressed using $U(N)$ generators. We
will use the notation that $k_i\times
k_j=k_i^{\mu}\theta_{\mu\nu}k_j^{\nu}$, then from Feynman rules we
know that structure constant $f^{ABC}$ and symmetric tensor
$d^{ABC}$ appear as\cite{Armoni:2000xr}
\bea f^{ABC}\cos(-{1\over 2}k_A\times k_B)-d^{ABC}\sin(-{1\over
2}k_A\times k_B)~ \eea
in each three-vertex. Using (\ref{unalgebra}) we can rewrite
$f^{ABC}$ and $d^{ABC}$ as trace of $U(N)$ generators, i.e., we have
\bea \cos{(-{1\over 2}k_A\times
k_B)}~\mbox{tr}([t^A,t^B]t^C)+i\sin{(-{1\over 2}k_A\times
k_B)}~\mbox{tr}(\{t^A,t^B\}t^C)~~~~\eea
assigned to each three-gluon vertex. Using Euler's formula $e^{i
x}=\cos x+i\sin x$ we can reorganize the former result as
\bea e^{-{i\over 2}k_A\times k_B}\mbox{tr}(t^At^Bt^C)-e^{-{i\over
2}k_B\times k_A} \mbox{tr}(t^Bt^At^C)~.~~~~\eea
This differs from result of conventional Yang-Mills theory with a
phase factor. In double line notation, we can diagrammatically
express the color structure of three-gluon vertex as in Fig.
\ref{3pcolor}.
Similar difference exists in four-gluon vertex, where from Feynman
rules we can see that for each such four-gluon vertex the color
generators appear as\cite{Armoni:2000xr}
\bean \mbox{tr}&&\Big[\left(-i\cos(-{1\over 2}k_A\times
k_B)[t^A,t^B]+\sin(-{1\over 2}k_A\times k_B)\{t^A,t^B\}\right)\\
&&\times\left(-i\cos(-{1\over 2}k_C\times
k_D)[t^C,t^D]+\sin(-{1\over 2}k_C\times
k_D)\{t^C,t^D\}\right)\Big]~.~~~\eean
With the same trick we can reorganize this expression as
\begin{eqnarray} &&-e^{-{i\over 2}(k_A\times k_B+k_C\times
k_D)}\mbox{tr}(t^At^Bt^Ct^D)+e^{-{i\over 2}(k_A\times k_B-k_C\times
k_D)}\mbox{tr}(t^At^Bt^Dt^C)\nonumber\\
&&+e^{{i\over 2}(k_A\times k_B-k_C\times
k_D)}\mbox{tr}(t^Bt^At^Ct^D)-e^{{i\over 2}(k_A\times k_B+k_C\times
k_D)}\mbox{tr}(t^Bt^At^Dt^C)~,~~~~\end{eqnarray}
and we can further use momentum conservation $k_D=-(k_A+k_B+k_C)$
for the first and third phase factors and $k_C=-(k_A+k_B+k_D)$ for
the second and fourth phase factors to express them in a more
systematic way. Combined with the property that $\theta^{\mu\nu}$ is
antisymmetric, we have the following result
\begin{eqnarray} &&-e^{-{i\over 2}(k_A\times k_B+k_A\times
k_C+k_B\times k_C)}\mbox{tr}(t^At^Bt^Ct^D)+e^{-{i\over 2}(k_A\times
k_B+k_A\times
k_D+k_B\times k_D)}\mbox{tr}(t^At^Bt^Dt^C)\nonumber\\
&&+e^{-{i\over 2}(k_B\times k_A+k_B\times k_C+k_A\times
k_C)}\mbox{tr}(t^Bt^At^Ct^D)-e^{-{i\over 2}(k_B\times k_A+k_B\times
k_D+k_A\times k_D)}\mbox{tr}(t^Bt^At^Dt^C)~.~~~~\end{eqnarray}
We see again that phase factors arise before color-ordered trace
strings and the value of phase depends on the color order. In double
line notation this color structure of four-gluon vertex can be
expressed as in Fig. \ref{4pcolor}.

There is another important relation about $t^A$ strings of $U(N)$
group. While some strings of $t^A$ are terminated by fundamental
indices, we have
\bea
(t^A)^{~\bar{\j}_1}_{i_1}(t^A)^{~\bar{\j}_2}_{i_2}=\delta^{~\bar{\j}_2}_{i_1}\delta^{~\bar{\j}_1}_{i_2}~.~~~\eea
This relation enables us to express tree-level amplitudes of many
gluons, which are constructed by three-gluon and four-gluon
vertices, into sum of color-ordered amplitudes.

%

\subsection{Color decomposition of noncommutative tree amplitudes}

All these above observations lead to the consequence that one can
write down the $n$-point tree amplitude as sum of single trace
terms, i.e., color-ordered amplitudes, with additional phase
factors. This is known as {\em color decomposition} of
noncommutative scattering amplitudes. More explicitly, we
have\cite{Berends:1987cv,Mangano:1987xk,Mangano:1988kk,Dixon:1996wi,Raju:2009yx}
\bea A^{tree}_{NC}(\{k_i,\lambda_i,a_i\})=g^{n-2}\sum_{\sigma\in
S_n/Z_n}\mbox{tr}(t^{A_{\sigma(a)}}\dots
t^{A_{\sigma(n)}})A^{tree}_{NC}(\sigma(1^{\lambda_1}),\ldots,\sigma(n^{\lambda_n}))~,~~~~\eea
where we have assumed that coupling constants of various vertices
containing $SU(N)$ gluons and $U(1)$ gluons are the same and equal
to $g$. $k_i$ and $\lambda_i$ are the gluon momenta and helicities,
and $A_{NC}^{tree}$ are partial amplitudes of noncommutative $U(N)$
Yang-Mills theory which contain all the kinematic information and
phase factors. $S_n$ is the set of all permutations of $n$ particles
and $Z_n$ is the subset of cyclic permutations, which preserves the
trace and should be eluded from summation in case of over counting.
$A_{NC}^{tree}$ can be expressed explicitly
as\cite{Filk:1996dm,Minwalla:1999px,Martin:1999aq,Hayakawa:1999zf,Hayakawa:1999yt,Matusis:2000jf,Douglas:2001ba}
\bea
A_{NC}(k_1,\ldots,k_n)=A_C(k_1,\ldots,k_n)\phi(k_1,\ldots,k_n)~,~~~\label{nc2c}\eea
with phase factor
\bea \phi(k_1,\ldots,k_n)=\exp\left[-{i\over 2}\sum_{1\leq
i<j<n}k_i\times k_j\right]~.~~~\eea
Since we will discuss only tree-level amplitudes in this note, the
superscript of "tree" has been thrown away for simplicity. The
multiplication in phase factor is defined as $k_i\times
k_j=k_i^{\mu}\theta_{\mu\nu}k_j^{\nu}$, and $A_{C}$ is color-ordered
amplitude of conventional Yang-Mills theory.

\subsection{Cyclic permutation and reflection of color order}

Tree amplitudes of conventional field theory possess some nice
properties which can be deduced directly from their color structure,
such as color cyclic relation and reflection relation. In
noncommutative $U(N)$ Yang-Mills theory, these relations might not
be trivially held, since we should also consider the effect of phase
factor after cyclic permutation or reflection. Let us first discuss
cyclic permutation. The phase factor is invariant under cyclic
permutation of gluons\cite{Douglas:2001ba}, which can be seen
directly from its definition, i.e., if we relabel $k_i$ as $k_{i+1}$
and $k_n$ as $k_1$, then
\bea \sum_{1\leq i<j<n}k_i\times k_j \to \sum_{2\leq i<j\leq
n}k_i\times k_j=\sum_{2\leq i<j<n}k_i\times k_j+\sum_{2\leq
i<n}k_i\times k_n~,~~~\label{phasecyclic}\eea
since $\sum_{2\leq i<n}k_i=-(k_1+k_n)$ and $k_n=-\sum_{1\leq
j<n}k_j$, and note that $k_i\times k_i=0$, the second term of
(\ref{phasecyclic}) is just $\sum_{1<j<n}k_1\times k_j$. After
combining this with the first term we reproduced the original phase
factor, thus proved the cyclic property of phase factor. In this
case, noncommutative tree amplitudes follow the same cyclic relation
as amplitudes of conventional Yang-Mills theory, i.e.,
\bea
A_{NC}(k_1,k_2,\ldots,k_n)=A_{NC}(k_n,k_1,\ldots,k_{n-1})~.~~~\label{cyclic}\eea

Next let us discuss the color reflection relation of noncommutative
Yang-Mills theory. When the color order is reversed, from the
definition of phase factor we have
\bea \phi(k_n,\ldots,k_1)=\exp\left[-{i\over 2}\sum_{n\geq
i>j>1}k_i\times k_j\right]=\exp\left[{i\over 2}\sum_{1<j<i\leq
n}k_j\times k_i\right]~,~~~\eea
where in the second step we used the antisymmetric property of
$\theta^{\mu\nu}$, so that $k_i\times k_j=-k_j\times k_i$. The
result is nothing but $\phi^{-1}(k_2,\ldots,k_n,k_1)$, which we have
shown before. Then use the property that phase factor is invariant
under cyclic permutation we get
\bea \phi(k_n,\ldots,k_1)=\phi^{-1}(k_1,\ldots,k_n)~.~~~~\eea
This is the color reflection relation of phase factor. Using
relation (\ref{nc2c}) and the known color reflection relation of
conventional Yang-Mills theory, we can deduce the color reflection
relation of noncommutative $U(N)$ Yang-Mills theory, i.e.,
\bea
A_{NC}(k_n,\ldots,k_1)=(-1)^nA_{NC}(k_1,\ldots,k_n)\phi^{-2}(k_1,\ldots,k_n)~.~~~\label{colorref}\eea
%

\section{BCFW recursion relation}

Since any amplitudes can be decomposed into color-ordered
amplitudes, we could only focus on these color-ordered amplitudes.
Is there an efficient method to compute these amplitudes? Or is
there a way to analyze these color-ordered amplitudes without
knowing their explicit expressions? We know that BCFW recursion
relation can be served as a good answer to these problems. In this
section we will try to extend BCFW recursion relation to
noncommutative $U(N)$ Yang-Mills theory.

\subsection{BCFW recursion relation for noncommutative $U(N)$ Yang-Mills theory}

\begin{figure}[t]
\center
  \includegraphics[width=6in]{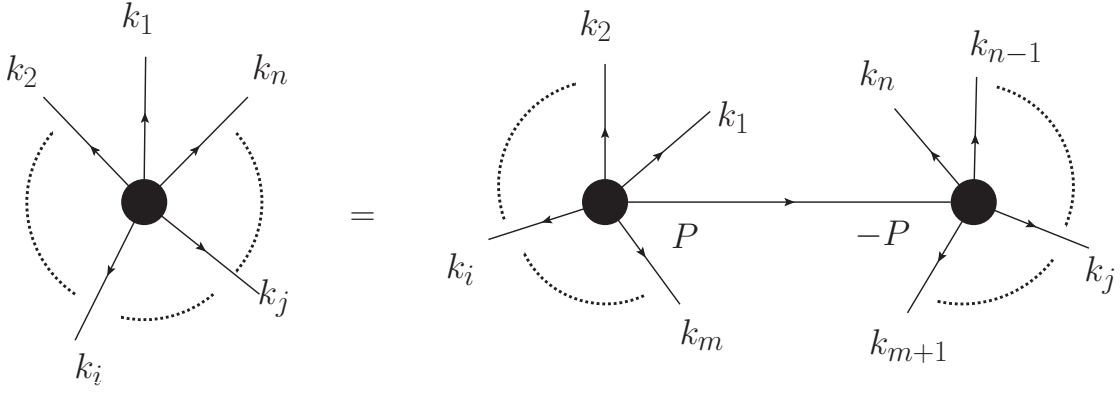}\\
  \caption{Phase factor of one single amplitude equals to the product of phase
  factors of two sub-amplitudes. }\label{phi}
\end{figure}

We have already seen from (\ref{nc2c}) that tree-level amplitudes of
noncommutative Yang-Mills theory differ from amplitudes of
conventional Yang-Mills theory in additional phase
factors\cite{Filk:1996dm,Minwalla:1999px,Martin:1999aq,Hayakawa:1999zf,Hayakawa:1999yt,Matusis:2000jf,Douglas:2001ba},
so when discussing the validation of BCFW recursion relation of
noncommutative Yang-Mills theory, we should focus on the discussion
of phase factor. Let us first consider the following property of
phase factor\cite{Filk:1996dm}
\bea
\phi(k_1,\ldots,k_n)=\phi(k_1,\ldots,k_m,P)\phi(-P,k_{m+1},\ldots,k_n)~,~~~\label{phi2subphi}\eea
which is nothing but the relation of phase factors shown in Fig.
\ref{phi}. In other words, the product of two phase factors of
sub-amplitudes, which are connected by a propagator, equals to the
phase factor of a single amplitude of the left hand side of figure
\ref{phi}.
All external momenta are going outward so that momentum conservation
takes the form $\sum_{1\leq i\leq n}k_i=0$, and the momentum of
propagator is $P=-\sum_{1\leq i\leq m}k_i$. To prove this property,
let us write down the phase factor as
\bean \phi(k_1,\ldots,k_n)&=&\exp\left[-{i\over 2}\sum_{1\leq
i<j<n}k_i\times k_j\right]=\exp\left[-{i\over 2}\sum_{1\leq
i<j\leq n}k_i\times k_j\right]\\
&=&\exp\left[-{i\over 2}\sum_{1\leq i<j\leq m}k_i\times k_j-{i\over
2}\sum_{\substack{1\leq i\leq m\\ m+1\leq j\leq n}}k_i\times
k_j-{i\over 2}\sum_{m+1\leq i<j\leq n}k_i\times k_j
\right]~.~~~\eean
The first line is true because $\sum_{1\leq i<n}k_i\times
k_n=-k_n\times k_n=0$. Let us deal with those three terms in the
second line. The first term is $\phi(k_1,\ldots,k_m,P)$ by
definition. The second term is zero since $\sum_{1\leq i\leq
m}k_i=-P$ and $\sum_{m+1\leq j\leq n}k_j=P$, so the summation equals
to $-P\times P=0$. The third term is $\phi(k_{m+1},\ldots,k_n,-P)$
by definition, and because of cyclic relation of phase factor this
is just $\phi(-P,k_{m+1},\ldots,k_n)$. This proved relation
(\ref{phi2subphi}). Note that we use only momentum conservation and
antisymmetric property of $\theta^{\mu\nu}$ through the proof, and
BCFW deformation of two selected momenta will always keep momentum
conversation, so the above argument is also held when momenta are
shifted.

Then let us discuss BCFW shifting of $n$-point noncommutative
amplitude. We pick up two momenta $k_i, k_j$ and take following
shifting
\bea \hat{k}_i=k_i-z\tilde{q}~,~~~\hat{k}_j=k_j+z\tilde{q}~.~~~\eea
Amplitudes of conventional Yang-Mills theory are often written down
in spinor formalism so that they have more compact and elegant form
(see \cite{Dixon:1996wi} and references there in). In spinor
formalism $\tilde{q}$ is often taken as
$\tilde{q}_{a\dot{a}}=\lambda_{ia}\tilde{\lambda}_{j\dot{a}}$, so
that $\tilde{q}^2=\tilde{q}\cdot k_i=\tilde{q}\cdot k_j=0$, and
$\hat{k}_i, \hat{k}_j$ are on-shell. We can always change
$\tilde{q}_{a\dot{a}}$ back to $\tilde{q}^{\mu}$ with $\sigma^{\mu
a\dot{a}}$ matrix. Tree amplitudes of noncommutative Yang-Mills
theory contain two parts, one is phase factor and the other is the
amplitude of conventional field theory. Since here we want to
discuss the effect of phase factor after BCFW shifting, we will
assume that for amplitude of conventional Yang-Mills theory this
shifting always leads to correct boundary
condition\cite{ArkaniHamed:2008yf,Benincasa:2007qj,Feng:2009ei,Feng:2010ku}.
Then after taking $(k_i, k_j)$-shifting the phase factor becomes
\bea
\phi(k_1,\ldots,\hat{k}_i,\ldots,\hat{k}_j,\ldots,k_n)=\phi(k_1,\ldots,k_n)\varphi(z)~,~~~\label{shiftphi}\eea
where
\bea \varphi(z)=\exp\left[-z{i\over
2}\left(k_i+2\sum_{i<l<j}k_l+k_j\right)\times
\tilde{q}\right]~.~~~\label{phiz}\eea
We see that phase factor is not invariant after $(k_i,k_j)$-shifting
and $z$-dependence enters into phase factor through $\varphi(z)$. It
is obviously that $\varphi(z)$ would not equal to one if $z\neq 0$,
since $(k_i+2\sum_{i<l<j}k_l+k_j)\times \tilde{q}$ does not
necessary vanish. In fact, result of
$(k_i+2\sum_{i<l<j}k_l+k_j)\times \tilde{q}$ depends not only on the
way of shifting but also on the value of $\theta^{\mu\nu}$. Is there
a suitably chosen $\tilde{q}$ that satisfies both requirements of
BCFW deformation and the vanishing of above result? It seems not
possible in 4-dimension space-time. $\tilde{q}$ is chosen that
$\tilde{q}\cdot k_i=\tilde{q}\cdot k_j=\tilde{q}^2=0$, and these
requirements are satisfied only when auxiliary momentum $\tilde{q}$
is complex\cite{ArkaniHamed:2008yf}. More explicitly, since $k_i$
and $k_j$ are massless, we could choose a suitable frame so that
$k_i=(E,E,0,0), k_j=(E,-E,0,0)$, and assume
$\tilde{q}=(E_q,q_x,q_y,q_z)$. Two equations $k_i\cdot
\tilde{q}=k_j\cdot \tilde{q}=0$ will determine two components of
$\tilde{q}$, and in the chosen frame we have $E_q=q_x=0$. The
massless condition of $\tilde{q}$ gives one more constraint on
$\tilde{q}$ and we have $q_y^2+q_z^2=0$. Then requirement of
$(k_i+2\sum_{i<l<j}k_l+k_j)\times \tilde{q}=0$ adds one more linear
constraint on $\tilde{q}$ so that $aq_y+bq_z=0$, and $a,b$ are
certain real constants. The solution of these two constraints is
$q_y=q_z=0$. In this case auxiliary momentum $\tilde{q}$ is just a
null vector, so we see that there are no non-trivial solution of
$\tilde{q}$ that satisfies all these requirements.

What is the problem if phase factor after shifting is not equal to
phase factor that are not being shifted? Generally speaking, we
would expect that BCFW recursion relation of noncommutative $U(N)$
Yang-Mills theory takes the same form as conventional Yang-Mills
theory, i.e.,
\bea
A_{NC}(k_1,\ldots,k_n)=\sum_{P_{ab}}A_{NC}^{h}(\ldots,\hat{k}_i(z_{ab}),\ldots,\hat{P}_{ab}(z_{ab})){1\over
P_{ab}^2}A_{NC}^{-h}(-\hat{P}_{ab}(z_{ab})\ldots,\hat{k}_j(z_{ab}),\ldots)~,~~~\label{notbcfw}\eea
where summation is over all possible helicities and propagators, and
$z_{ab}$ is the solution of $\hat{P}^2_{ab}(z)=0$. This is obviously
not true from (\ref{nc2c}). The phase factor for left hand side of
(\ref{notbcfw}) is $\phi(k_1,\ldots,k_n)$. Each term of BCFW
expansion in right hand side of (\ref{notbcfw}) has a phase factor
$\phi(k_1,\ldots,\hat{k}_i(z_{ab}),\ldots,\hat{k}_j(z_{ab}),\ldots,k_n)$,
after using cyclic relation of phase factor and (\ref{phi2subphi}).
So the phase factor in the left hand side of (\ref{notbcfw}) does
not equal to phase factors in the right hand side of
(\ref{notbcfw}), and even phase factors in the right hand side do
not equal to each other.

In order to write down a suitable BCFW recursion relation for
noncommutative $U(N)$ Yang-Mills theory, let us recall the
supersymmetric BCFW recursion
relation\cite{ArkaniHamed:2008gz,Bianchi:2008pu,Brandhuber:2008pf,Elvang:2008na}.
We are forced to take $(k_i,k_j)$-shifting so that energy-momentum
conservation is satisfied after shifting, and in supersymmetric BCFW
recursion relation we are forced to take one more Grassmann variable
$\eta$-shifting so that super-energy-momentum conservation is
satisfied. In noncommutative case, the one that must be kept
invariant is phase factor, so it is reasonable to consider taking
some kind of phase-deformation. It is also natural to think this
problem from the view of singularities. In tree amplitudes of
noncommutative $U(N)$ Yang-Mills theory, there are so called
essential singularities from phase factors and ordinary
singularities from propagators. We could eliminate essential
singularities by multiplying one additional phase factor, so that
there are only ordinary singularities from propagators, and BCFW
recursion relation is valid. More specifically, we have
\begin{eqnarray}
A_{NC}(k_1,\ldots,k_n)\phi^{-1}(k_1,\ldots,k_n)&=&\sum_{P_{ab}}\Big[
A_{NC}^{h}(\ldots,\hat{k}_i(z_{ab}),\ldots,\hat{P}_{ab}(z_{ab}))\phi^{-1}(\ldots,\hat{k}_i(z_{ab}),\ldots,\hat{P}_{ab}(z_{ab}))\nonumber\\
&&\times{1\over
P_{ab}^2}A_{NC}^{-h}(-\hat{P}_{ab}(z_{ab})\ldots,\hat{k}_j(z_{ab}),\ldots)\phi^{-1}(-\hat{P}_{ab}(z_{ab})\ldots,\hat{k}_j(z_{ab}),\ldots)\Big]~.~~~\nonumber\\\end{eqnarray}
The above recursion relation can be further simplified by
multiplying $\phi(k_1,\ldots,k_n)$ in both left hand side and right
hand side. Using (\ref{phi2subphi}) and cyclic relation of phase
factor we have
\bea
\phi^{-1}(\ldots,\hat{k}_i(z_{ab}),\ldots,\hat{P}_{ab}(z_{ab}))\phi^{-1}(-\hat{P}_{ab}(z_{ab})\ldots,\hat{k}_j(z_{ab}),\ldots)=\phi^{-1}(k_1,\ldots,\hat{k}_i(z_{ab}),\ldots,\hat{k}_j(z_{ab}),\ldots,k_n)~,~~~\eea
and after using (\ref{shiftphi}) we have
\bea
\phi(k_1,\ldots,k_n)\phi^{-1}(k_1,\ldots,\hat{k}_i(z_{ab}),\ldots,\hat{k}_j(z_{ab}),\ldots,k_n)=\varphi^{-1}(z_{ab})~,~~~\eea
where $\varphi(z)$ is defined in (\ref{phiz}). Then we have
\bea A_{NC}(k_1,\ldots,k_n)=\sum_{P_{ab}}
A_{NC}^{h}(\ldots,\hat{k}_i(z_{ab}),\ldots,\hat{P}_{ab}(z_{ab})){\varphi^{-1}(z_{ab})\over
P_{ab}^2}A_{NC}^{-h}(-\hat{P}_{ab}(z_{ab})\ldots,\hat{k}_j(z_{ab}),\ldots)~.~~~\label{bcfw}\eea

Since essential singularities have been removed from amplitudes of
noncommutative Yang-Mills theory in BCFW recursion relation
(\ref{bcfw}), there would be no problems considering singularities
or large $z$ behavior of $A_{NC}$ in complex plane as long as BCFW
recursion relation of conventional field theory is held. Of course
(\ref{bcfw}) is not the best way to compute tree amplitudes of
noncommutative $U(N)$ Yang-Mills theory. Thanks to relation
(\ref{nc2c}), in order to get amplitudes of noncommutative
Yang-Mills theory, we can simply obtain tree amplitudes of
conventional field theory by ordinary BCFW recursion relation and
multiply them with corresponding phase factors. But recursion
relation (\ref{bcfw}) could be served as a good tool for
noncommutative theory in S-Matrix program where, for example, one
may want to prove some identities of amplitudes without knowing
explicit expressions of these amplitudes, etc.

\subsection{Three-point amplitudes}
The on-shell BCFW recursion relation allows us to construct any tree
amplitudes from three-point amplitudes theoretically. After writing
down BCFW recursion relation for noncommutative Yang-Mills theory,
we further want to get noncommutative three-point amplitudes.
Three-point amplitudes are fundamental amplitudes, but we know that
there are no three-point scattering amplitudes when all three
momenta are
real\cite{Benincasa:2007xk,ArkaniHamed:2008gz,Schuster:2008nh,He:2008nj}.
In order to satisfy energy-momentum conservation and massless
conditions, we have $p_i\cdot p_j=0$ for any momenta. In spinor
formalism these equations are $\Spaa{i~j}\Spbb{i~j}=0$, which means
that either $\Spaa{i~j}=0$ or $\Spbb{i~j}=0$ if complex momenta are
allowed, so the three-point amplitudes are purely holomorphic or
anti-homomorphic. In conventional field theory it is known that for
particles of spin $s$, the amplitudes of helicity configurations
$(+,+,-)$ and $(-,-,+)$ should be of the form
\bea A^{--+}=\left({\Spaa{1~2}^4\over
\Spaa{1~2}\Spaa{2~3}\Spaa{3~1}}\right)^s~,~~~~A^{++-}=\left({\Spbb{1~2}^4\over
\Spbb{1~2}\Spbb{2~3}\Spbb{3~1}}\right)^s~.~~~~\label{3point}\eea
These forms are determined up to an overall dimensionless coupling
constant\cite{ArkaniHamed:2008gz}. To construct three-point
amplitudes of noncommutative field theory we could naively add one
additional phase factor on (\ref{3point}). For noncommutative $U(N)$
Yang-Mills theory we conclude that
\bea A_{NC}(1^-,2^-,3^+)=e^{-{i\over 2}k_1\times
k_2}\left({\Spaa{1~2}^4\over
\Spaa{1~2}\Spaa{2~3}\Spaa{3~1}}\right)~,~~~~A_{NC}(1^+,2^+,3^-)=e^{-{i\over
2}k_1\times k_2}\left({\Spbb{1~2}^4\over
\Spbb{1~2}\Spbb{2~3}\Spbb{3~1}}\right)~.~~~\label{nc3point}\eea
Due to energy-momentum conservation and antisymmetry of
$\theta^{\mu\nu}$ we have $k_1\times k_2=k_2\times k_3=k_3\times
k_1$, thus the phase factor can be written in other equivalent
forms. Different from conventional field theory, we have an
additional parameter $\theta^{\mu\nu}$ in noncommutative Yang-Mills
theory and this parameter enables us to construct dimensionless
constant from momenta. For example the multiplication $k_i\times
k_j$ is dimensionless, and phase factors (\ref{nc3point}) are also
dimensionless. Then forms of (\ref{nc3point}) are determined up to
overall coupling constant and dimensionless constants constructed
from $\theta^{\mu\nu}$. It would be an interesting work to discuss
consistency conditions on the S-matrix of noncommutative field
theory.

There is one property of three-point amplitudes we want to address.
By direct verification we see that
\bea
A_{NC}(k_1,k_2,k_3)=-A_{NC}(k_3,k_2,k_1)\phi^{-2}(k_3,k_2,k_1)~.~~~\label{3pbasic}\eea
This relation can be considered as a special case of color
reflection relation (\ref{colorref}) when $n=3$. It is simple to
prove (\ref{colorref}) by BCFW recursion relation of noncommutative
Yang-Mills theory starting from three-point relation
(\ref{3pbasic}), and (\ref{3pbasic}) will play an important role in
the later demonstration of amplitude relations of noncommutative
Yang-Mills theory.

\subsection{A simple example: computing four-point amplitude $A_{NC}(1^+,2^+,3^-,4^-)$}

As a verification of these three-point amplitudes and noncommutative
BCFW recursion relation, a simple example will be given below to
show how to construct four-point amplitude from three-point
amplitudes. Let us consider four-point amplitude of helicity
configuration $(++--)$. We will take $(k_1,k_4)$-shifting, i.e.,
\bea \hat{k}_1=k_1-z\tilde{q}~,~~~\hat{k}_4=k_4+z\tilde{q}~,~~~\eea
and in the spinor formalism we have $\tilde{q}=|4\rangle|1]$.
Consequently we have shifted the holomorphic part of $k_1$ and the
anti-holomorphic part of $k_4$. According to (\ref{bcfw}) we have
\bea
A_{NC}(1^+,2^+,3^-,4^-)=A_{NC}(\hat{1}^+,2^+,-\hat{P}^-){\exp[z_{14}{i\over
2}(k_1+2(k_2+k_3)+k_4)\times \tilde{q}]\over
s_{12}}A_{NC}(\hat{P}^+,3^-,\hat{4}^-)~,~~~\label{4from3}\eea
where $\hat{P}=\hat{k}_1+k_2$ and $z_{14}$ is the solution of
$\hat{P}^2=0$, and $s_{12}=(k_1+k_2)^2$ is the propagator. In order
to get $z_{14}$ we should solve equation
\bea \hat{P}^2=\Spaa{1~2}\Spbb{1~2}-z\Spaa{2~4}\Spbb{2~1}=0~,~~~\eea
and the result is
\bea z_{14}=-{\Spaa{1~2}\over \Spaa{2~4}}~.~~~\eea
The three-point sub-amplitudes can be written down directly as
\bea A_{NC}(\hat{1}^+,2^+,-\hat{P}^-)=e^{-{i\over 2}\hat{k}_1\times
k_2}{\Spbb{1~2}^4\over
\Spbb{1~2}[2~\hat{P}][\hat{P}~1]}~,~~~~A_{NC}(\hat{P}^+,3^-,\hat{4}^-)=e^{-{i\over
2}k_3\times \hat{k}_4}{\Spaa{3~4}^4\over
\langle\hat{P}~3\rangle\Spaa{3~4}\langle 4~\hat{P}\rangle}~.~~~\eea
Firstly let us compute the phase factor in right hand side of
(\ref{4from3}), which is
\bean &&\exp\left[-{i\over 2}\left(k_1\times
k_2-z_{14}\tilde{q}\times
k_2-z_{14}(k_1+2(k_2+k_3)+k_4)\times\tilde{q}+k_3\times
k_4+z_{14}k_3\times\tilde{q}\right)\right]\\
&=&\exp\left[-{i\over 2}(k_1\times k_2+k_3\times
k_4-z_{14}(k_1+k_2+k_3+k_4)\times \tilde{q})\right]\\
&=&\exp\left[-{i\over 2}(k_1\times k_2+k_1\times k_3+k_2\times
k_3)\right]\equiv\phi(1,2,3,4)~,~~~\eean
where we have used $k_4=-(k_1+k_2+k_3)$ and antisymmetry of
$\theta^{\mu\nu}$ in the second step. Then we want to compute
\bean {\Spbb{1~2}^4\over \Spbb{1~2}[2~\hat{P}][\hat{P}~1]}{1\over
s_{12}}{\Spaa{3~4}^4\over \langle\hat{P}~3\rangle\Spaa{3~4}\langle
4~\hat{P}\rangle} &=&{\Spbb{1~2}^3\Spaa{3~4}^3\over
s_{12}[2|\hat{P}|3\rangle[1|\hat{P}|4\rangle}~.~~~\eean
Since $\hat{P}=|1\rangle|1]+|2\rangle|2]-z_{14}|4\rangle|1]$ it is
easy to see that
\bean [1|\hat{P}|4\rangle&=&\Spba{1|2|4}~,~~~\eean
and
\bean
[2|\hat{P}|3\rangle&=&\Spba{2|1|3}-z_{14}\Spbb{2~1}\Spaa{4~3}\\
&=&{\Spbb{2~1}\over
\Spaa{2~4}}\left(\Spaa{2~4}\Spaa{1~3}+\Spaa{1~2}\Spaa{4~3}\right)\\
&=&{\Spbb{2~1}\Spaa{2~3}\Spaa{1~4}\over \Spaa{2~4}}~,~~~\eean
where we have used Schouten identity in the second step. By writing
$s_{12}=\Spaa{1~2}\Spbb{1~2}$ and put all the above results back to
(\ref{4from3}) we get the final result
\bea A_{NC}(1^+,2^+,3^-,4^-)={\Spaa{3~4}^4\over
\Spaa{1~2}\Spaa{2~3}\Spaa{3~4}\Spaa{4~1}}\phi(1,2,3,4)~.~~~\eea
This result is same as the one obtained from (\ref{nc2c}).

\section{Relations of amplitudes in noncommutative $U(N)$ Yang-Mills theory}

We know that any tree amplitudes can be expressed as sum of
color-ordered amplitudes, and there is so called BCFW recursion
relation which can be used to efficiently compute or analyze these
amplitudes. But thanks to some nontrivial relations of amplitudes we
do not need to calculate all these color-ordered amplitudes. In
noncommutative $U(N)$ Yang-Mills theory, there are also analogies of
these nontrivial relations. In this section, we will write down the
analogies of $U(1)$-decoupling, KK and BCJ relations in
noncommutative Yang-Mills theory and prove them with BCFW recursion
relation of noncommutative version recursively, starting from the
verified three-point relations.

\subsection{Modified $U(1)$-decoupling relation}
In conventional $U(N)$ Yang-Mills theory, $U(1)$ gauge boson is
identified as photon, and $SU(N)$ as gluons. Since any amplitude
containing extra $U(1)$ photon must vanish and $U(1)$ generator
commutes with other generators, we could get $U(1)$-decoupling
relation. But in noncommutative $U(N)$ Yang-Mills theory there are
interactions among $U(1)$ gluons and $SU(N)$
gluons\cite{Armoni:2000xr}, so amplitude containing $U(1)$ gluons
exists, and we could not get similar $U(1)$-decoupling relation of
$A_{NC}$ by simply replacing $A_C$ with $A_{NC}$ in ordinary
$U(1)$-decoupling relation. So a modified $U(1)$-decoupling relation
is necessary for tree amplitudes of noncommutative Yang-Mills
theory. It is reasonable to infer that phase factor would appear in
noncommutative $U(1)$-decoupling relation, since phase factor is a
character of noncommutative Yang-Mills theory. In fact, deduced from
(\ref{nc2c}) we could get the noncommutative $U(1)$-decoupling
relation
\bea
\sum_{\sigma\in~cyclic}A_{NC}(k_1,\sigma(k_2,\ldots,k_n))\phi^{-1}(k_1,\sigma(k_2,\ldots,k_n))=0~,~~~\label{u1}\eea
where $\sigma$ is the set of cyclic permutation of
$(k_2,\ldots,k_n)$. The appearance of phase factor $\phi^{-1}$ is an
effect of noncommutative space-time structure. In the limit that
$\theta^{\mu\nu}\to 0$ the space-time turns to be normal, and
$\phi^{-1}\to 1$ while $A_{NC}\to A_{C}$, thus noncommutative
$U(1)$-decoupling relation (\ref{u1}) turns to ordinary
$U(1)$-decoupling relation of gauge theory.

The $U(1)$-decoupling relation of three-point is
\bea
A_{NC}(k_1,k_2,k_3)\phi^{-1}(k_1,k_2,k_3)+A_{NC}(k_1,k_3,k_2)\phi^{-1}(k_1,k_3,k_2)=0~.~~~\eea
Through cyclic relation (\ref{cyclic}) we have
$A_{NC}(k_1,k_3,k_2)=A_{NC}(k_3,k_2,k_1)$, and using the color
reflection relation of phase factor, i.e.,
$\phi(k_1,k_2,k_3)=\phi^{-1}(k_3,k_2,k_1)$, we can rewrite the above
relation as
\bea
A_{NC}(k_1,k_2,k_3)=-A_{NC}(k_3,k_2,k_1)\phi^{-2}(k_3,k_2,k_1)~.~~~\eea
This is the three-point relation (\ref{3pbasic}) we have verified.
We want to prove $U(1)$-decoupling relation of general point
recursively through BCFW recursion relation of noncommutative
Yang-Mills theory (\ref{bcfw}). Suppose noncommutative
$U(1)$-decoupling relation is satisfied by all noncommutative tree
amplitudes less than $n$-point, we will demonstrate that it is true
for $n$-point tree amplitude.

We can rewrite the $n$-point noncommutative $U(1)$-decoupling
relation as
 \bea
 \sum_{\sigma\in
 cyclic}A_{NC}(1,\sigma(2,3,\ldots,n))\phi^{-1}(1,\sigma(2,3,\ldots,n))=0~.~~~~\label{n-pu1}
 \eea
For expression simplicity we introduce split notation $|$, which is
defined as
 \bea
 A_{NC}(\hat{1},\ldots,i|i+1,\ldots,\hat{n})\equiv\sum_{h}A_{NC}(\hat{1},\ldots,i,\hat{P}_{1i}^h){1\over
 P_{1i}^2}A_{NC}(-\hat{P}^{-h}_{1i},i+1,\ldots,\hat{n})~.~~~~\label{split}
 \eea
In this notation, we expand every amplitude in (\ref{n-pu1}) using
BCFW expansion under $(n-1,n)$-shifting, for example
 \bea
 A_{NC}(1,2,\ldots,n-1,n)=\sum_{i=1}^{n-3}A_{NC}(\hat{n},1,\ldots,i|i+1,\ldots,n-2,\widehat{n-1})\varphi^{-1}(z_{n1\cdots i})~,~~~
 \eea
where $\varphi(z)$ is
 \bea
 \varphi(z)=\exp\Big[z{i\over 2}(k_n+k_{n-1})\times\tilde{q}\Big]~,~~~
 \eea
and $z_{n1\cdots i}$ is the solution of
$\hat{P}^2=(\hat{k}_n+k_1+\cdots+k_i)^2=0$. Note that under
different manners of shifting $z_{n1\cdots i}$ will take different
values. For example, we can take $\Spab{n-1|n}$-shifting or
$\Spab{n|n-1}$-shifting under $(n-1,n)$-shifting, and $z_{n1\cdots
i}$ is different under these two manners of shifting. But this
property will not affect the proof, since this proof is general and
does not depend on the details of BCFW shifting.

We also have
 \bea
 A_{NC}(1,n-1,n,2,\ldots,n-3,n-2)=\sum_{i=2}^{n-2}A_{NC}(\hat{n},2,\ldots,i|i+1,\ldots,n-2,1,\widehat{n-1})\varphi^{-1}(z_{n2\cdots i})~,~~~
 \eea
where $z_{n2\cdots i}$ is determined by the equation
$(\hat{k}_n+k_2+k_3+\cdots+k_i)^2=0$. Another two special cases are
 \bea
 A_{NC}(1,n,2,\ldots,n-2,n-1)&=&\sum_{i=2}^{n-2}A_{NC}(\hat{n},2,\ldots,i|i+1,\ldots,n-2,\widehat{n-1},1)\varphi^{-1}(z_{n2\cdots i})\nonumber\\
 &+&\sum_{i=2}^{n-2}A_{NC}(1,\hat{n},2,3,\ldots,i-1|i,\ldots,n-2,\widehat{n-1})\varphi^{-1}(z_{n1\cdots i-1})~,~~~\eea
and
 \bea
 A_{NC}(1,3,4,\ldots,n-1,n,2)&=&A_{NC}(\hat{n},2|1,3,\ldots,n-2,\widehat{n-1})\varphi^{-1}(z_{n2})\nonumber\\
 &+&\sum_{i=3}^{n-2}A_{NC}(\hat{n},2,1,3,\ldots,i-1|i,\ldots,n-2,\widehat{n-1})\varphi^{-1}(z_{n1\cdots i-1})~.~~~
 \eea
For other general cases $(4\leqslant j\leqslant n-2)$, we have
 \bea\label{generalcase}
 A_{NC}(1,j,\ldots,n,2,\ldots,j-1)&=&\sum_{i=2}^{j-1}A_{NC}(\hat{n},2,3,\ldots,i|i+1,\ldots,j-1,1,j,\ldots,n-2,\widehat{n-1})\varphi^{-1}(z_{n2\cdots i})\nonumber\\
 &+&\sum_{i=j}^{n-2}A_{NC}(\hat{n},2,\ldots,j-1,1,j,\ldots,i-1|i,\ldots,\widehat{n-1})\varphi^{-1}(z_{n1\cdots i}).
 \eea
In the above expansions, we have ignored phase factors in
(\ref{n-pu1}). Let us take them back and consider the general case
(\ref{generalcase}). According to (\ref{shiftphi}), we have
 \bea\label{takebackpf}
&&\varphi^{-1}(z_{n2\cdots i})\phi^{-1}(1,j,\ldots,n,2,\ldots,j-1)\nonumber\\
&=&\varphi^{-1}(z_{n2\cdots i})
\phi^{-1}(n,2,3,\ldots,i,i+1,\ldots,j-1,1,j,\ldots,n-2,n-1)\nonumber\\
&=&\phi^{-1}(\hat{n}(z_{n2\cdots i}),2,3,\ldots,i,i+1,\ldots,j-1,1,j,\ldots,n-2,\widehat{n-1}(z_{n2\cdots i}))\nonumber\\
&=&\phi^{-1}(\hat{n}(z_{n2\cdots i}),2,3,\ldots,i,\hat{P}_{n2\cdots
i})\phi^{-1}(-\hat{P}_{n2\cdots i},
i+1,\ldots,j-1,1,j,\ldots,n-2,\widehat{n-1}(z_{n2\cdots i}))~.~~~
 \eea
Similar operation can be applied to $\varphi^{-1}(z_{n1\cdots i})$
and phases in the special cases. We see that all the expansions can
be divided into two groups: one contains parameter $z_{n2\cdots i}$,
and the other contains $z_{n1\cdots i}$. Firstly, let us consider
all terms containing $z_{n2\cdots i}$, and we can arrange them as
follows,
 \bea
&~&A_{NC}(\hat{n},2|1,3,\ldots,n-2,\widehat{n-1})\varphi^{-1}(z_{n2})\phi^{-1}(n,2,1,3,\ldots,n-1)\nonumber\\
&+&\sum_{j=4}^{n-2}\sum_{i=2}^{j-1}A_{NC}(\hat{n},2,3,\ldots,i|i+1,\ldots,j-1,1,j,\ldots,n-2,\widehat{n-1})\varphi^{-1}(z_{n2\cdots i})\phi^{-1}(n,2,\ldots,j,1,j+1,\ldots,n-1)\nonumber\\
&+&\sum_{i=2}^{n-2}A_{NC}(\hat{n},2,\ldots,i|i+1,\ldots,n-2,1,\widehat{n-1})\varphi^{-1}(z_{n2\cdots i})\phi^{-1}(n,2,\ldots,n-2,1,n-1)\nonumber\\
&+&\sum_{i=2}^{n-2}A_{NC}(\hat{n},2,\ldots,i|i+1,\ldots,n-2,\widehat{n-1},1)\varphi^{-1}(z_{n2\cdots
i})\phi^{-1}(n,2,\ldots,n-1,1)~.~~~
 \eea
Then according to (\ref{takebackpf}) we can obtain terms as
 \bea
 &&\sum_h \Big[A_{NC}(\hat{n},2,\ldots,i,\hat{P}^h_{n2\cdots i})\phi^{-1}(\hat{n}(z_{n2\cdots i}),2,3,\ldots,i,\hat{P}^h_{n2\cdots i})
 {1\over P_{n2\cdots i}^2}\nonumber\\
 &&\times\sum_{\sigma\in cyclic}A_{NC}(-\hat{P}^{-h}_{n2\cdots i}, \sigma(1, i+1,\ldots, n-2, \widehat{n-1}))
 \phi^{-1}(-\hat{P}^{-h}_{n2\cdots i}, \sigma(1, i+1,\ldots, n-2,
 \widehat{n-1}))\Big]~.~~~
 \eea
Using $(n-i+1)$-point noncommutative $U(1)$-decoupling relation, we
see the above summation vanishes. So all terms containing
$z_{n2\cdots i}$ are zero. The same argument holds for all terms
containing $z_{n1\cdots i}$. Thus we proved the noncommutative
$U(1)$-decoupling relation for $n$-point tree amplitudes.

\subsection{Noncommutative KK relations}
General KK relations should also be modified to fit the requirement
of noncommutative Yang-Mills theory, and we conclude that
noncommutative KK relations take the form
 \bea  A_{NC}(1,\{\a\}, n,\{\b\})\phi^{-1}(1,\{\a\}, n,\{\b\})
= (-1)^{n_\b}\sum_{\sigma\in OP(\{\a\},\{\b^T\})} A_{NC}(1,\sigma,
n)\phi^{-1}(1,\sigma, n)~.~~~~\label{KK}
 \eea
where $OP(\{\a\},\{\b^T\})$ is the ordered permutations between sets
$\{\a\}$ and $\{\b^T\}$, i.e., the permutation keeps order of set
$\{\a\}$ and set $\{\b^T\}$. $\{\b^T\}$ is the reverse ordering of
$\{\b\}$.

Let us consider a special case that $\{\a\}$ is empty. The KK
relations become
 \bea
  A_{NC}(1, n, \{\b\})\phi^{-1}(1, n, \{\b\})
&=& (-1)^{n_\b} A_{NC}(1, \{\b^T\}, n)\phi^{-1}(1, \{\b^T\}, n)\nonumber\\
&=&(-1)^{n_\b} A_{NC}(\{\b^T\}, n, 1)\phi^{-1}(\{\b^T\}, n, 1)~.~~~
 \eea
Using color reflection relation for phase factor, we have
 \bea
\phi^{-1}(1, n, \{\b\})=\phi( \{\b^T\}, n, 1)~.~~~
 \eea
Then we get $(n_\beta +2)$-point color reflection relation
 \bea
 A_{NC}(1, n, \{\b\})=(-1)^{n_\b} A_{NC}(\{\b^T\}, n, 1)\phi^{-2}(\{\b^T\}, n, 1).
 \eea
Thus we see that noncommutative color reflection relation is a
special case of noncommutative KK relations.

Then let us consider another special case when $\{\b\}$ has only one
element. In this case we have
 \bea
  A_{NC}(1, \{\a\}, n, \b)\phi^{-1}(1, \{\a\}, n, \b)
 = (-1)\sum_{OP} A_{NC}(1, OP(\{\a\}\cup\b), n)\phi^{-1}(1, OP(\{\a\}\cup\b),
 n)~.~~~
 \eea
Since $A_{NC}$ and phase factors are invariant under cyclic
permutations, the above equation become
 \bea
 &&A_{NC}(1, \{\a\}, n, \b)\phi^{-1}(1, \{\a\}, n, \b)
 +\sum_{OP}A_{NC}(1, OP(\{\a\}\cup\b), n)\phi^{-1}(1, OP(\{\a\}\cup\b),
 n)\nonumber\\
 &=&\sum_{\sigma\in cyclic}A_{NC}(\b, \sigma(1, \{\a\}, n))\phi^{-1}(\b, \sigma(1, \{\a\},
 n))~.~~~
 \eea
This is nothing but noncommutative $U(1)$-decoupling relation. It
can be easily checked that the special case that $\{\a\}$ containing
only one element is also noncommutative $U(1)$-decoupling relation.

After checking above several special cases, we use noncommutative
BCFW recursion relation to prove general noncommutative KK
relations, which is
 \bea\label{kkgeneral}
&&A_{NC}(1,\{\a_1, \ldots, \a_l\}, n,\{\b_1,\ldots,
\b_m\})\phi^{-1}(1,\{\a_1, \ldots, \a_l\}, n,\{\b_1,\ldots,
\b_m\})\nonumber\\
&=& (-1)^m \sum_{\sigma\in OP(\{\a\},\{\b^T\})} A_{NC}(1,\sigma,
n)\phi^{-1}(1,\sigma, n)~.~~~
 \eea
We will use noncommutative BCFW recursion relation to expand tree
amplitudes in both sides of KK relations. It can be shown that each
term in the left hand side belongs to the right hand side and both
sides have the same number of terms. For the right hand side of
noncommutative KK relations, using BCFW recursion relation and
$(n,1)$-shifting, we have
 \bea\label{rhexp}
&&(-1)^m \sum_{\sigma\in OP(\{\a\},\{\b^T\})} A_{NC}(1,\sigma,
n)\phi^{-1}(1,\sigma, n)\nonumber\\
&=&(-1)^m \sum_{\sigma\in OP(\{\a\},\{\b^T\})}\sum_{h;i=1}^{l+m-1}
A_{NC}(\hat{1},\sigma_1,\ldots,\sigma_i|\sigma_{i+1},\ldots,\sigma_{l+m},
\hat{n})\phi^{-1}(\hat{1},\sigma, \hat{n})~.~~~
 \eea
The total number of terms in the above expansion is
  \bea\label{totalnumberright}
  \frac{(l+m)!}{l!m!}(l+m-1)~.~~~
  \eea
For the left hand side of noncommutative KK relations, we would
firstly deal with amplitudes of noncommutative Yang-Mills theory and
phase factors separately, then combine them to get the final
results. For amplitudes on the left hand side, we have
 \bea\label{nckk1}
&&A_{NC}(1,\{\a_1, \ldots, \a_l\}, n,\{\b_1,\ldots, \b_m\})\nonumber\\
&=&\sum_{h;j=0}^{m-1}A_{NC}(\{\b_{j+1},\ldots, \b_m\}, \hat{1},
\hat{P}_{lj}^h){1\over P_{lj}^2}A_{NC}(-\hat{P}_{lj}^{-h}, \{\a_1,
\ldots, \a_l\}, \hat{n}, \{\b_1, \ldots, \b_j\})\varphi^{-1}(z_{lj})\nonumber\\
&+&\sum_{h;j=1}^{m}A_{NC}(\{\b_{j+1},\ldots, \b_m\}, \hat{1},
\{\a_1, \ldots, \a_l\}, \hat{P}_{0j}^h){1\over
P_{0j}^2}A_{NC}(-\hat{P}_{0j}^{-h},
 \hat{n}, \{\b_1, \ldots, \b_j\})\varphi^{-1}(z_{0j})\nonumber\\
 &+&\sum_{i=1}^{l-1}\sum_{h;j=0}^{m}\Big[A_{NC}(\{\b_{j+1},\ldots, \b_m\},
 \hat{1},\{\a_1, \ldots, \a_i\}, \hat{P}_{i,j}^h){1\over P_{i,j}^2}\nonumber\\
 &~~~&~~~~~~~~~~~~~~\times A_{NC}(-\hat{P}_{i,j}^{-h}, \{\a_{i+1},
\ldots, \a_l\}, \hat{n}, \{\b_1, \ldots,
\b_j\})\varphi^{-1}(z_{i,j})\Big]~,~~~
 \eea
where we have $P_{lj}\equiv P_{n\alpha_1\cdots \alpha_l\beta_1\cdots
\beta_j},z_{lj}\equiv z_{n\alpha_1\cdots \alpha_l\beta_1\cdots
\beta_j}$ in the second line, $P_{0j}\equiv P_{n\beta_1\cdots
\beta_j},z_{0j}\equiv z_{n\beta_1\cdots \beta_j}$ in the third line
and $P_{i,j}\equiv P_{n\alpha_{i+1}\cdots \alpha_l\beta_1\cdots
\beta_j},z_{i,j}\equiv z_{n\alpha_{i+1}\cdots \alpha_l\beta_1\cdots
\beta_j}$ in the fourth line. Then we discuss phase factors in this
expansion. Multiplying each term in the expansion with
 \bea
 \phi^{-1}(1,\{\a_1, \ldots, \a_l\}, n,\{\b_1,\ldots, \b_m\})~,~~~
 \eea
and using relation (\ref{shiftphi}) and cyclic symmetry of phase
factors, we have
 \begin{eqnarray}
&~&\phi^{-1}(1,\{\a_1, \ldots, \a_l\}, n,\{\b_1,\ldots,
\b_m\})\varphi^{-1}(z_{lj})\nonumber\\
&=&\phi^{-1}(\hat{1}(z_{lj}),\{\a_1,
\ldots,
\a_l\}, \hat{n}(z_{lj}),\{\b_1,\ldots, \b_m\})\nonumber\\
&=&\phi^{-1}(\{\b_{j+1},\ldots, \b_m\}, \hat{1},\{\a_1, \ldots,
\a_l\}, \hat{n}, \{\b_1, \ldots, \b_j\})\nonumber\\
&=&\phi^{-1}(\{\b_{j+1},\ldots, \b_m\}, \hat{1},
\hat{P}_{lj}^h)\phi^{-1}(-\hat{P}_{lj}^{-h}, \{\a_1, \ldots, \a_l\},
\hat{n}, \{\b_1, \ldots, \b_j\})~,~~~\label{phase1}\end{eqnarray}
and
\begin{eqnarray}
 &~&\phi^{-1}(1,\{\a_1,
\ldots, \a_l\}, n,\{\b_1,\ldots,
\b_m\})\varphi^{-1}(z_{0j})\nonumber\\
&=&\phi^{-1}(\{\b_{j+1},\ldots, \b_m\}, \hat{1}, \{\a_1, \ldots,
\a_l\}, \hat{P}_{0j}^h)\phi^{-1}(-\hat{P}_{0j}^{-h},
 \hat{n}, \{\b_1, \ldots, \b_j\})~,~~~~\label{phase2}\end{eqnarray}
and
\begin{eqnarray}
 &~&\phi^{-1}(1,\{\a_1,
\ldots, \a_l\}, n,\{\b_1,\ldots, \b_m\})\varphi^{-1}(z_{i,j})\nonumber\\
&=&\phi^{-1}(\{\b_{j+1},\ldots, \b_m\},
 \hat{1},\{\a_1, \ldots, \a_i\}, \hat{P}_{i,j}^h)\phi^{-1}(-\hat{P}_{i,j}^{-h}, \{\a_{i+1},
\ldots, \a_l\}, \hat{n}, \{\b_1, \ldots,
\b_j\})\varphi^{-1}(z_{i,j})~.~~~\label{phase3}\end{eqnarray}
We can add the phase factor (\ref{phase1}) back to the first term in
the right hand side of equation (\ref{nckk1}) and get
 \bea\label{kk1total0}
&&\sum A_{NC}(\{\b_{j+1},\ldots, \b_m\}, \hat{1},
\hat{P}_{lj}^h){1\over P_{lj}^2}A_{NC}(-\hat{P}_{lj}^{-h}, \{\a_1,
\ldots, \a_l\}, \hat{n},
\{\b_1, \ldots, \b_j\})\nonumber\\
&\times&\phi^{-1}(\{\b_{j+1},\ldots, \b_m\}, \hat{1},
\hat{P}_{lj}^h)\phi^{-1}(-\hat{P}_{lj}^{-h}, \{\a_1, \ldots, \a_l\},
\hat{n}, \{\b_1, \ldots, \b_j\})~.~~~
 \eea
Using $(m-j+2)$-point and $(l+j+2)$-point noncommutative KK
relations, we have
 \bea
&&A_{NC}(\{\b_{j+1},\ldots, \b_m\}, \hat{1},
\hat{P}_{lj}^h)\phi^{-1}(\{\b_{j+1},\ldots, \b_m\}, \hat{1},
\hat{P}_{lj}^h)\nonumber\\&&=(-1)^{m-j}A_{NC}(\hat{1},\{\b_{m},\ldots,
\b_{j+1}\}, \hat{P}_{lj}^h)\phi^{-1}(\hat{1},\{\b_{m},\ldots,
\b_{j+1}\}, \hat{P}_{lj}^h)~,~~~\eea
and
\bea &&A_{NC}(-\hat{P}_{lj}^{-h}, \{\a_1, \ldots, \a_l\}, \hat{n},
\{\b_1, \ldots, \b_j\})\phi^{-1}(-\hat{P}_{lj}^{-h}, \{\a_1, \ldots,
\a_l\}, \hat{n}, \{\b_1, \ldots,
\b_j\})\nonumber\\&&=(-1)^j\sum_{\sigma\in OP(\{\a_1, \ldots,
\a_l\}, \{\b_j, \ldots, \b_1\} )}A_{NC}(-\hat{P}_{lj}^{-h}, \sigma,
\hat{n})\phi^{-1}(-\hat{P}_{lj}^{-h}, \sigma, \hat{n})~.~~~ \eea
So the finial result of equation (\ref{kk1total0}) is
 \bea
 (-1)^m\sum_{h;j=0}^{m-1}\sum_{\sigma\in OP(\{\a\}, \{\b_j, \ldots, \b_1\} )}A_{NC}(\hat{1},\{\b_{m},\ldots,
\b_{j+1}\}, \hat{P}_{lj}^h){1\over
P_{lj}^2}A_{NC}(-\hat{P}_{lj}^{-h}, \sigma,
\hat{n})\phi^{-1}(\hat{1},\{\b_{m},\ldots, \b_{j+1}\},\sigma,
\hat{n})~.~~~\nonumber\\
 \eea
It is easy to check that these terms belong to the right hand side
of noncommutative KK relations (\ref{rhexp}). Using the same trick,
it is easy to show that the last two terms on the right hand side of
equation (\ref{nckk1}) with their phase factors (\ref{phase2}) and
(\ref{phase3}) also belong to the right hand side of noncommutative
KK relations (\ref{rhexp}). So in order to complete this proof we
want to show that the number of terms on both sides are equal. The
total number of terms on the left hand side is
 \bea
 \sum_{j=0}^{m-1}C_{l+j}^l+\sum_{j=1}^{m}C_{l+m-j}^l+\sum_{i=1}^{l-1}\sum_{j=0}^m
 C_{m+i-j}^i C_{l+j-i}^j~.~~~
 \eea
This number is the same as the total number (\ref{totalnumberright})
on the right hand side of noncommutative KK relations, which can be
checked in Mathematica.

\subsection{Noncommutative BCJ relations}
Finally let us discuss BCJ relations of amplitudes of noncommutative
$U(N)$ Yang-Mills theory. In conventional gauge theory, it is argued
that general BCJ relations can be derived from level one BCJ
relations, combined by KK and $U(1)$-decoupling
relations\cite{Feng:2010my,Jia:2010nz}. One can expect this is also
true in noncommutative Yang-Mills theory. In order to prove general
noncommutative BCJ relations, we just need to prove level one
noncommutative BCJ relations. As we have done before, we will prove
these relations recursively. The four-point BCJ relations can be
checked directly, and if we suppose all less than $n$-point
noncommutative BCJ relations are correct, then we could prove that
$n$-point noncommutative BCJ relations are also correct. The general
expression of level one noncommutative Yang-Mills theory is
 \bea
 \sum_{i=3}^n
\Big[A_{NC}(1,3,\cdots,i-1,2,i,\cdots,n)\phi^{-1}(1,3,\cdots,i-1,2,i,\cdots,n)\sum_{j=i}^n
s_{2j}\Big]=0~.~~~\label{BCJ}
 \eea
All noncommutative tree amplitudes in the above summation can be
written down explicitly as
 \bea
 &&A_{NC}(1, 2, 3, 4, \dots, n-1, n)~,~~~\nonumber\\
 &&A_{NC}(1, 3, 2, 4, \dots, n-1, n )~,~~~\nonumber\\
 &&\ldots~,~~~\nonumber\\
 &&A_{NC}(1, 3, 4, \dots, 2, n-1, n)~,~~~\nonumber\\
  &&A_{NC}(1, 3, 4, \dots, n-1, 2, n)~.~~~
 \eea
Using noncommutative BCFW recursion relation under $(n,1)$-shifting,
all expansion terms in BCJ relations can be arranged as follows,
 \bea\label{bcjexpansion1}
 &&\sum_{i=5}^n\sum_{k=3}^{i-2}A_{NC}(\hat{1}, 3, \ldots, k|k+1,
 \ldots,i-1,2,i,\ldots,\hat{n})\phi^{-1}(\sum_{j=i}^{\hat{n}}s_{2j}+s_{2n}-s_{2\hat{n}})\\\label{bcjexpansion2}
 &+&\sum_{k=3}^{n-1}A_{NC}(\hat{1}, 3, \ldots, k|2, k+1, \ldots,
 \hat{n})\phi^{-1}(\sum_{j=k+1}^{\hat{n}}s_{2j}+s_{2n}-s_{2\hat{n}})\\\label{bcjexpansion3}
 &+&\sum_{i=3}^{n-2}\sum_{k=i}^{n-2}A_{NC}(\hat{1}, 3,
 \ldots,i-1,2,i, \ldots, k|k+1,\ldots,\hat{n})\phi^{-1}(\sum_{j=i}^{\hat{n}}s_{2j}+s_{2n}-s_{2\hat{n}})\\\label{bcjexpansion4}
 &+&\sum_{k=2}^{n-2}A_{NC}(\hat{1}, 3,\ldots,
 k,2|k+1,\ldots,\hat{n})\phi^{-1}(\sum_{j=k+1}^{\hat{n}}s_{2j}+s_{2n}-s_{2\hat{n}})~.~~~
 \eea
In the above expression, for simplicity we do not write down phase
factors explicitly, but one should remind that momentum ordering of
each phase factor is the same as their corresponding amplitude. One
should also remind that $\phi^{-1}$ comes from multiplication of two
different phase factors: one phase factor $\varphi^{-1}(z)$ from
noncommutative BCFW recursion relation and the other phase factor
from the modified BCJ relations (\ref{BCJ}) with all momenta
unshifted, thus according to (\ref{shiftphi}) we know that momenta
in $\phi^{-1}$ are shifted. In this subsection we will always take
this abbreviation.

Note that momenta in sub-amplitudes are shifted, in order to use
lower point BCJ relations, we should also shift momenta in the
kinetic factors, i.e., we should replace $s_{2n}$ with
 $s_{2\hat{n}}+s_{2n}-s_{2\hat{n}}$. Now let us consider the sum of
 first two terms in above equation. Summation $\sum_{i=5}^n\sum_{k=3}^{i-2}$ in equation
(\ref{bcjexpansion1}) can be replaced by
$\sum_{k=3}^{n-2}\sum_{i=k+2}^n$ after changing the summation order.
For a given $k$, terms involving $\sum s_{2j}$ can be calculated as
 \bea
 &&\sum_{i=k+2}^n \sum_h A_{NC}(\hat{1}, 3,\ldots,k,\hat{P}_k^h)\phi^{-1}{1\over
 P_k^2}A_{NC}(-\hat{P}_k^{-h},k+1,\ldots,i-1,2,i,\ldots,\hat{n})\phi^{-1}\sum_{j=i}^{\hat{n}}s_{2j}\nonumber\\
&+&\sum_h A_{NC}(\hat{1}, 3,\ldots,k,\hat{P}_k^h)\phi^{-1}{1\over
 P_k^2}A_{NC}(-\hat{P}_k^{-h},2,k+1,\ldots,
 \hat{n})\phi^{-1}\sum_{j=k+1}^{\hat{n}}s_{2j}~,~~~
 \eea
where we should emphasize again that $\phi^{-1}$ are just
abbreviations and momentum ordering of these phase factors are the
same as those amplitudes just before them, and momenta are shifted.
Using $(n-k+2)$-point noncommutative BCJ relations, it is easy to
see that sum of the above equation is zero. Then let us consider
terms involving $s_{2n}-s_{2\hat{n}}$, which are
 \bea
 &&\sum_{k=3}^{n-2}\Big[\sum_{i=k+2}^n \sum_h A_{NC}(\hat{1},\ldots,k,\hat{P}_k^h)\phi^{-1}{1\over
 P_k^2}A_{NC}(-\hat{P}_k^{-h},k+1,\ldots,i-1,2,i,\ldots,\hat{n})\phi^{-1}(s_{2n}-s_{2\hat{n}})\nonumber\\
 &+&\sum_h A_{NC}(\hat{1}, 3,\ldots,k,\hat{P}_k^h)\phi^{-1}{1\over
 P_k^2}A_{NC}(-\hat{P}_k^{-h},2,k+1,\ldots,
 \hat{n})\phi^{-1}(s_{2n}-s_{2\hat{n}})\Big]\nonumber\\
&+&\sum_h A_{NC}(\hat{1},
3,\ldots,n-1,\hat{P}_{2n}^h)\phi^{-1}{1\over
 P_{2n}^2}A_{NC}(-\hat{P}_{2n}^{-h},2,
 \hat{n})\phi^{-1}(s_{2n}-s_{2\hat{n}})~.~~~
 \eea
Using cyclic symmetry and $(n-k+2)$-point noncommutative
$U(1)$-decoupling relation, the above equation can be rewritten as
 \bea
&&-\sum_{k=3}^{n-2}\sum_h A_{NC}(\hat{1},
3,\ldots,k,\hat{P}_k^h)\phi^{-1}{1\over
 P_k^2}A_{NC}(-\hat{P}_k^{-h},k+1,\ldots,
 \hat{n}, 2)\phi^{-1}(s_{2n}-s_{2\hat{n}})\nonumber\\
 &&-\sum_h A_{NC}(\hat{1},
3,\ldots,n-1,\hat{P}_{2n}^h)\phi^{-1}{1\over
 P_{2n}^2}A_{NC}(-\hat{P}_{2n}^{-h}, \hat{n},2)\phi^{-1}(s_{2n}-s_{2\hat{n}})\nonumber\\
 &=&-\sum_{k=3}^{n-1}\sum_h A_{NC}(\hat{1},
3,\ldots,k,\hat{P}_k^h)\phi^{-1}{1\over
 P_k^2}A_{NC}(-\hat{P}_k^{-h},k+1,\ldots,
 \hat{n}, 2)\phi^{-1}(s_{2n}-s_{2\hat{n}})~.~~~
 \eea
This is the final result of equation (\ref{bcjexpansion1}) plus
(\ref{bcjexpansion2}). Using the same trick, we could systematically
perform calculation of the last two equations (\ref{bcjexpansion3})
and (\ref{bcjexpansion4}), which yields the result
 \bea
 -\sum_{k=3}^{n-1}\sum_h A_{NC}(2,\hat{1},
3,\ldots,k-1,\hat{P}_k^h)\phi^{-1}{1\over
 P_k^2}A_{NC}(-\hat{P}_k^{-h},k,\ldots,
 \hat{n})\phi^{-1}(s_{2n}-s_{2\hat{n}}).
 \eea
Then the total sum of BCFW expansions of BCJ relations is
 \bea\label{bcjfinal}
-\sum_{k=3}^{n-1}\sum_h &&\Big[A_{NC}(\hat{1},
3,\ldots,k,\hat{P}_k^h)\phi^{-1}{1\over
 P_k^2}A_{NC}(-\hat{P}_k^{-h},k+1,\ldots,
 \hat{n}, 2)\phi^{-1}(s_{2n}-s_{2\hat{n}})\nonumber\\
  &+&A_{NC}(2,\hat{1},
3,\ldots,k-1,\hat{P}_k^h)\phi^{-1}{1\over
 P_k^2}A_{NC}(-\hat{P}_k^{-h},k,\ldots,
 \hat{n})\phi^{-1}(s_{2n}-s_{2\hat{n}})\Big]~.~~~
 \eea
What is the meaning of this result? To understand this problem, let
us pay attention to phase factors. Phase factor for the first term
of (\ref{bcjfinal}) is
\bea \phi^{-1}(\hat{1},
3,\ldots,k,\hat{P}_k)\phi^{-1}(-\hat{P}_k,k+1,\ldots,
 \hat{n}, 2)=\phi^{-1}(\hat{1},3,\ldots,\hat{n},2)~,~~~\eea
and phase factor for the second term of (\ref{bcjfinal}) is
\bea \phi^{-1}(2,\hat{1},
3,\ldots,k-1,\hat{P}_k)\phi^{-1}(-\hat{P}_k,k,\ldots,
 \hat{n})=\phi^{-1}(2,\hat{1},3,\ldots,n-1,\hat{n})~.~~~\eea
Note that $z_k$ in above phase factors are determined by equations
of propagator $\hat{P}_k^2=0$. Then let us consider integration
\bea \oint {dz \over
z}A_{NC}(2,\hat{1},3,\ldots,n-1,\hat{n})\phi^{-1}(2,\hat{1},3,\ldots,n-1,\hat{n})s_{2\hat{n}}(z)~.~~~\label{closeint}\eea
Residue at pole $z=0$ yields
\bea A_{NC}(2,1,3,\ldots,n)\phi^{-1}(2,1,3,\ldots,n)s_{2n}~,~~~\eea
where we can further use BCFW recursion relation to expand $A_{NC}$
and get
 \bea
\sum_{k=3}^{n-1}\sum_h &&\Big[A_{NC}(\hat{1},
3,\ldots,k,\hat{P}_k^h){\phi^{-1}(z_k)\over
 P_k^2}A_{NC}(-\hat{P}_k^{-h},k+1,\ldots,
 \hat{n}, 2)\phi^{-1}(2,1,3,\ldots,n)s_{2n}\nonumber\\
  &+&A_{NC}(2,\hat{1},
3,\ldots,k-1,\hat{P}_k^h){\phi^{-1}(z_k)\over
 P_k^2}A_{NC}(-\hat{P}_k^{-h},k,\ldots,
 \hat{n})\phi^{-1}(2,1,3,\ldots,n)s_{2n}\Big]~.~~~
 \eea
Sum of residues at poles coming from propagators is
 \bea
-\sum_{k=3}^{n-1}\sum_h &&\Big[A_{NC}(\hat{1},
3,\ldots,k,\hat{P}_k^h){1\over
 P_k^2}A_{NC}(-\hat{P}_k^{-h},k+1,\ldots,
 \hat{n}, 2)\phi^{-1}(2,\hat{1},3,\ldots,n-1,\hat{n})s_{2\hat{n}}\nonumber\\
  &+&A_{NC}(2,\hat{1},
3,\ldots,k-1,\hat{P}_k^h){1\over
 P_k^2}A_{NC}(-\hat{P}_k^{-h},k,\ldots,
 \hat{n})\phi^{-1}(2,\hat{1},3,\ldots,n-1,\hat{n})s_{2\hat{n}}\Big]~.~~~
 \eea
So the final result of integration (\ref{closeint}) differs from
(\ref{bcjfinal}) in an overall minus sign. But we know that
\bea
A_{NC}(2,\hat{1},3,\ldots,n-1,\hat{n})\phi^{-1}(2,\hat{1},3,\ldots,n-1,\hat{n})=A_{C}(2,\hat{1},3,\ldots,n-1,\hat{n})~,~~~\eea
and because the shifted momenta $(1,n)$ are not adjacent in
$A_{C}(2,\hat{1},3,\ldots,n-1,\hat{n})$, this amplitude behaves as
$1/z^2$ in the boundary\cite{ArkaniHamed:2008gz}. Thus integration
(\ref{closeint}) equals to zero, and so is the result
(\ref{bcjfinal}). Thus we proved the noncommutative $n$-point BCJ
relations.

\section{Conclusion}

Following the idea of S-matrix program, in this note we proposed a
modified BCFW recursion relation for noncommutative $U(N)$
Yang-Mills theory. By using noncommutative BCFW recursion relation,
we proved noncommutative analogies of $U(1)$-decoupling, KK and BCJ
relations. We expect that similar modification to supersymmetric
BCFW recursion relation would also valid when supersymmetry is
considered. In \cite{Raju:2009yx} a basis for non-planar one-loop
amplitudes of noncommutative $U(N)$ Yang-Mills theory has been
proposed, it is also interesting to see if one can efficiently get
coefficients of one-loop amplitudes by using noncommutative BCFW
recursion relation. The study of BCFW recursion relation in nonlocal
field theories might also deepen our understanding of field theory,
thus it is of interest to extend on-shell recursion method to other
nonlocal theories.

\section*{Acknowledgments}
We would like to thank Prof. Bo Feng for useful discussions. This
work is supported by the Fundamental Research Funds for the Central
Universities with contract number 2009QNA3015.


\end{document}